\documentclass[twocolumn,showpacs,preprintnumbers,amsmath,amssymb,prl]{revtex4}

\usepackage{graphicx}
\usepackage{dcolumn}
\usepackage{bm}

\begin{document}

\title{The Near--Threshold Production of $\phi$ Mesons in $pp$ Collisions}

\author{
  M.\,Hartmann,$^1$
  Y.\,Maeda,$^{1,2}$ 
  I.\,Keshelashvilli,$^{1,3}$ 
  H.R.\,Koch,$^1$ 
  S.\,Mikirtytchiants,$^4$\\
  S.\,Barsov,$^4$ 
  W.\,Borgs,$^1$
  M.\,B\"uscher,$^1$
  V.I.\,Dimitrov,$^5$
  S.\,Dymov,$^6$
  V.\,Hejny,$^1$
  V.\,Kleber,$^7$
  V.\,Koptev,$^4$
  P.\,Kulessa,$^{1,8}$
  T.\,Mersmann,$^9$
  S.\,Merzliakov,$^6$
  A.\,Mussgiller,$^1$
  M.\,Nekipelov,$^1$
  M.\,Nioradze,$^3$
  H.\,Ohm,$^1$
  K.\,Pysz,$^8$
  R.\,Schleichert,$^1$
  H.J.\,Stein,$^1$
  H.\,Str\"oher,$^1$
  K.--H.\,Watzlawik,$^1$
  and P.\,W\"ustner$^{10}$
} \email{M.Hartmann@FZ-Juelich.de}

\affiliation{$^1$Institut f\"ur Kernphysik, Forschungszentrum
J\"ulich, 52425 J\"ulich, Germany}%
\affiliation{$^2$Institut f\"ur Kernphysik, Universit\"at zu
K\"oln, 50937 K\"oln, Germany}%
\affiliation{$^3$High Energy Physics Institute, Tbilisi State
University, 0186 Tbilisi, Georgia}%
\affiliation{$^4$High Energy Physics Department, Petersburg
Nuclear Physics Institute, 188350 Gatchina, Russia}%
\affiliation{$^5$Idaho Accelerator Center,
83209 Pocatello, USA}%
\affiliation{$^6$Laboratory of Nuclear Problems, Joint Institute 
for Nuclear Research, 141980 Dubna, Russia}%
\affiliation{$^7$Physikalisches Institut, Universit\"at 
Bonn, 53115 Bonn, Germany}%
\affiliation{$^8$H.~Niewodnicza\'{n}ski Institute of Nuclear
Physics PAN, 31342 Krak\'{o}w, Poland}%
\affiliation{$^9$Institut f\"ur Kernphysik, Universit\"at
M\"unster, 48149 M\"unster, Germany}%
\affiliation{$^{10}$Zentralinstitut f\"ur Elektronik,
Forschungszentrum J\"ulich, 52425 J\"ulich, Germany}

\date{\today}

\begin{abstract}
The $pp\,{\to}\,pp\phi$ reaction has been studied at the Cooler
Synchrotron COSY--J\"ulich, using the internal beam and ANKE
facility. Total cross sections have been determined at three
excess energies $\epsilon$ near the production threshold. The
differential cross section closest to threshold at
$\epsilon=18.5$\,MeV exhibits a clear $S$--wave dominance as well
as a noticeable effect due to the proton--proton final state
interaction. Taken together with data for $pp \omega$--production,
a significant enhancement of the $\phi/\omega$ ratio of a factor 8
is found compared to predictions based on the Okubo--Zweig--Iizuka
rule.
\end{abstract}


\maketitle

Meson production near threshold has the potential to clarify
important questions of hadron physics in the non-perturbative
regime of quantum chromodynamics due to its comparatively simple
scheme of interpretation. The production of light vector mesons,
$\rho(770)$, $\omega$(782) and $\phi$(1020), quark anti-quark
states with their spins aligned ($J^P{=}1^-$) and without open
strangeness, has been investigated with both hadronic and
electromagnetic probes in order to study production
mechanisms~\cite{TsushimaFaesslerKaptari}, coupling
constants~\cite{TsushimaFaesslerKaptari}, modifications in nuclear
medium~\cite{Naruki} and in particular the so--called
Okubo--Zweig--Iizuka (OZI) rule~\cite{OkuboZweigIizuka}. This rule
states that processes with disconnected quark lines between
initial and final states are suppressed compared to those where
the incident quarks continue through to the exit channel. As a
result, the production of ideally mixed $\phi$--mesons (quark
content $s\bar{s}$) in a reaction $A+B\,{\to}\,\phi X$ is reduced
compared to $A+B\,{\to}\,\omega X$ ($\omega$ is a linear
combination of $u\bar{u}+d\bar{d}$) under similar kinematical
conditions. Taking into account deviations from ideal mixing
between singlet and octet vector mesons, Lipkin predicted a ratio
of single $\phi$ to $\omega$ production of
$R_\mathrm{\phi{/}\omega}=4.2\,\times\!10^{-3}\equiv
R_\mathrm{OZI}$~\cite{Lipkin,PDG}. However, strong enhancements of the
experimental $R_\mathrm{\phi{/}\omega}$ compared to
$R_\mathrm{OZI}$ have been observed (an overview is given in
Ref.~\cite{Nomokonov}), in particular in $\bar{p}p$ annihilations,
where $R_\mathrm{\phi{/}\omega}$ can be as large as
$\sim100\!\times\!R_\mathrm{OZI}$~\cite{Amsler}. Here a strong
correlation of the $\phi$ meson yield with the spin--triplet
fraction of the initial state was found~\cite{Bertin}, and this in
part motivated the suggestion of a polarized internal strangeness
component in a polarized nucleon~\cite{Ellis}. However, alternative
explanations, such as two--step kaon--exchange
models~\cite{Meissner1,Locher}, have also been advanced. Since
vector--meson production in close--to--threshold $pp\,{\to}\,ppV$
reactions must proceed via the spin--triplet entrance channel, the
investigation of the cross section ratio
$\sigma(pp\,{\to}\,pp\phi)/\sigma(pp\,{\to}\,pp\omega)$ at small
excess energies $\epsilon$ should provide a clean way of
investigating possible violations of the OZI rule.

Total cross sections for $\omega$--production in proton--proton
collisions have been measured in a range of excess--energy
$\epsilon$ from a few MeV up to several
GeV~\cite{Hibou,Abd,LanBoe}, whereas data for $pp\phi$ are very
scarce. Two total cross sections of $\phi$ production have been
obtained for $\epsilon\sim\!(2-4)\,$GeV, but with rather limited
accuracy~\cite{Baldi,Blobel}. At low excess energy, a single
measurement of total and differential cross sections has been made
by the DISTO collaboration at
$\epsilon\!=\!83$\,MeV~\cite{Balestra}. In combination with the
$\omega$ cross section of COSY-TOF at
$\epsilon\!=\!92\,$MeV~\cite{Abd}, this yields
$R_\mathrm{\phi/\omega}\sim7\!\times\!R_\mathrm{OZI}$. The
differential distributions from DISTO indicate that $\phi$
production at that energy proceeds dominantly via the $^3P_1$
($pp$) entrance channel, though other partial waves do contribute
significantly. To clarify this, it is crucial to extend the
measurements to such small excess energies that only the lowest
partial waves can contribute. Such measurements have become
feasible at the internal proton beam of the Cooler Synchrotron
COSY at the Research Center J\"ulich, using the ANKE target and
detector facility. Here we report on the results for $\phi$
production in proton-proton collisions at three beam momenta,
corresponding to excess energies of $\epsilon\!=\!18.5$, 34.5 and
75.9$\,$MeV.

ANKE is a magnetic spectrometer~\cite{ANKE} situated at the
internal beam of COSY. It comprises three dipole
magnets D1---D3, which guide the circulating beam through a
variable chicane. The central C--shaped spectrometer dipole D2,
with a maximum field strength 1.6\,T, is placed downstream of the
target position. D2 is used to separate the reaction products from
the circulating beam, deflecting them towards charged--particle
detectors on the left/right side of the beam for negative/positive
charges. The hydrogen cluster--jet target used provided areal
densities of \mbox{$\sim
5\!\times\!10^{14}\,\textrm{cm}^{-2}$}~\cite{GasTarget}. The
average luminosity during the experiment was determined through
the simultaneous measurement of $pp$ elastic scattering. By
detecting one fast forward--going proton
($\vartheta\!=\!5.0^{\circ} - 8.5^{\circ}$) in appropriate
detectors, elastic events were easily separated from background.
Taking the corresponding cross sections from the SAID
database~\cite{ArndtMH}, luminosities between $(1.5-3.2) \times
10^{31}\textrm{cm}^{-2}\textrm{s}^{-1}$ were determined. The
uncertainties in these contribute $4\,\%$ (at
$\epsilon\!=\!18.5\,$MeV), $6\,\%$ (34.5\,MeV), and $9\,\%$
(75.9\,MeV), respectively, to the final systematic error in the
total cross sections.

The $pp\,{\to}\,pp\phi$ reaction has been studied by detecting the
$K^{+}K^{-}$ decay of $\phi$--mesons in coincidence with one of
the forward--going protons, requiring that the missing mass be
consistent with that of the non--observed second proton. Particle
identification relies on time--of--flight (TOF) measurements and
the determination of particle momenta. In the initial step,
positive kaons are selected by a procedure described in detail in
Ref.~\cite{ANKEKAON}, using TOF between START and STOP
scintillation counters of a dedicated $K^+$ detection system.
Secondly, both the coincident $K^{-}$ and forward--going
proton are selected from the time--of--flight differences between
the STOP counters --- in the negative as well as in the forward
detector system --- with respect to the positive STOP counter that was
hit by the $K^+$.
These two TOF selections, as well as the selection for the $K^+$, were done 
inside $\pm\,3\,\sigma$.
The absolute time calibration of all negative and forward STOP
counters in conjunction with all of the positive STOP counters was
performed using the abundant $\pi^{+}\pi^{-}$ and $\pi^{+}p$ pairs. The
final selection of the $pp\,{\to}\,ppK^{+}K^{-}$ reaction was made
by a $\pm\,3\,\sigma$ missing--mass cut on the non--detected proton. 
This leads to $400-1800$ identified $ppK^+K^-$ events
depending on the energy.
The estimated background inside the proton cut window is $5\,\%$ (at
$\epsilon\!=\!18.5\,$MeV), $12\,\%$ (34.5\,MeV) and $18\,\%$
(75.9\,MeV).

\begin{figure}[t]
  \vspace*{+1mm}
  \centering
  \includegraphics[clip,width=0.23\textwidth]{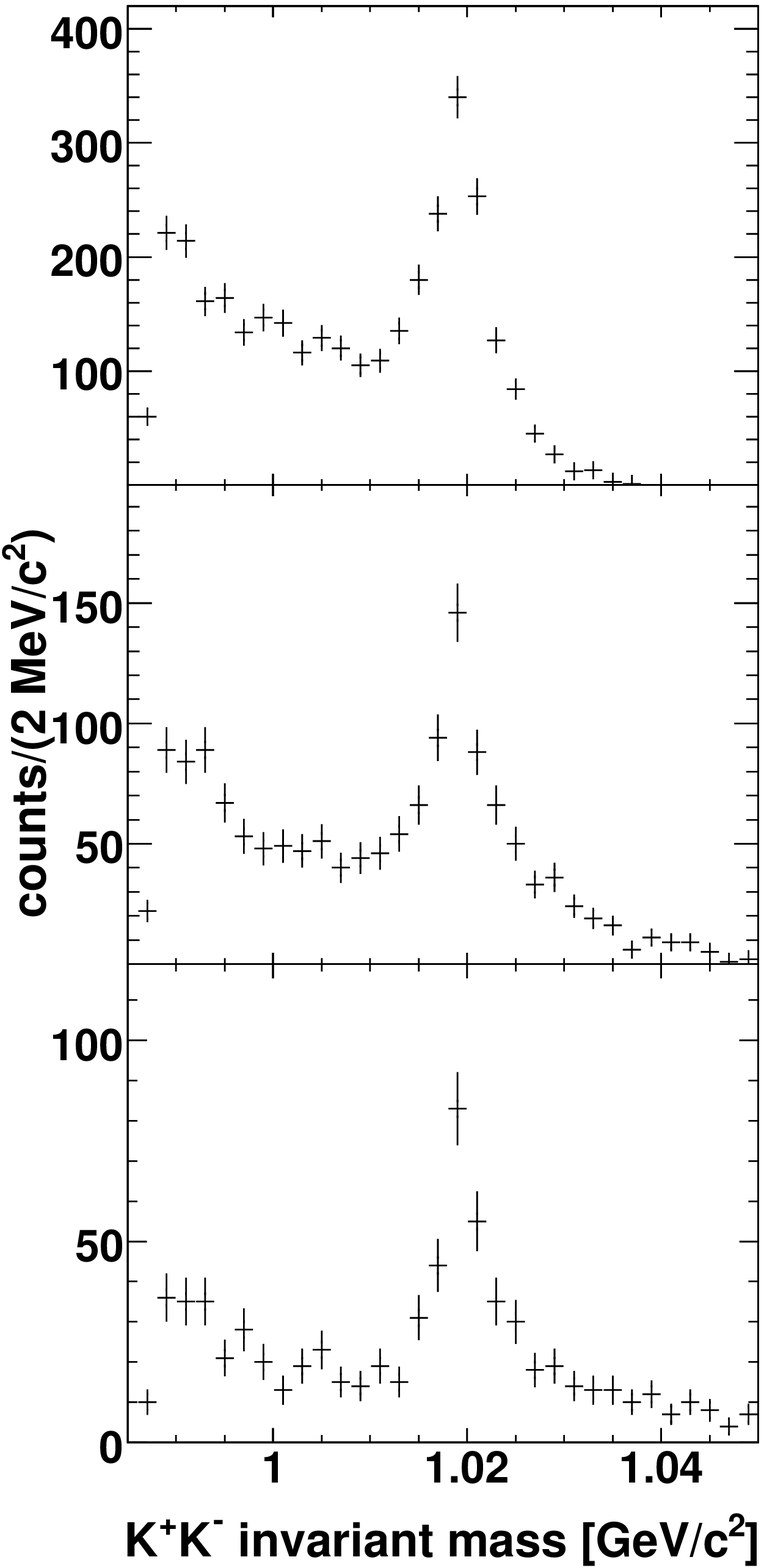}
  \includegraphics[clip,width=0.23\textwidth]{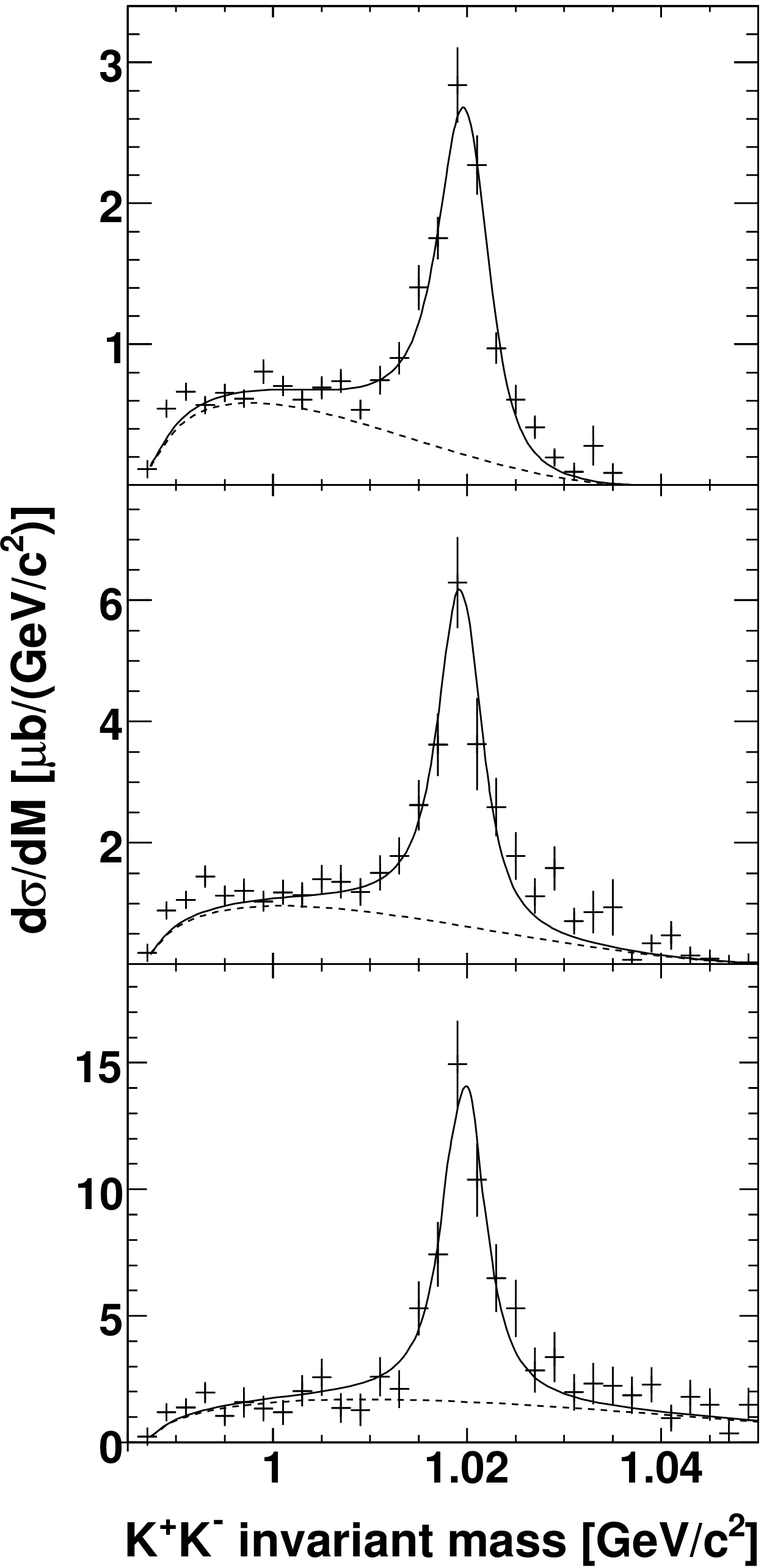}
  \vspace*{-2mm}
  \caption{
$K^{+}K^{-}$ invariant mass distributions at
$\epsilon=18.5\,$MeV (top), 34.5\,MeV (middle), and 75.9\,MeV
(bottom). Measured raw distributions are shown on the l.h.s. while
the corresponding cross sections d$\sigma$/d$M$ are plotted on the
right. Only statistical uncertainties are shown. The cross
sections contain a non-resonant part (dashed line, based on
four-body phase space) and a $\phi$-meson contribution. The solid
line is the sum of both and includes effects of the detector
resolution.
  }
  \label{IM}
\vspace*{-2mm}
\end{figure}

The left panel of Fig.~\ref{IM} shows the $K^+K^-$ invariant--mass
distributions in the region around $1\,$GeV/c$^2$. For all three
beam energies, a clean $\phi$ peak is observed at
$1.02\,$GeV/c$^2$ on top of a smooth background of non--resonant
kaon--pair production. The right panel shows the corresponding
$pp\,{\to}\,pp\phi\,{\to}\,ppK^+K^-$ differential cross sections,
\mbox{i.e.} distributions corrected for detector acceptances (see
below). Contributions from misidentified particles have been
subtracted using data from outside the proton peak in the
missing--mass distributions, adding $3\,\%$ (18.5\,MeV), $7\,\%$
(34.5\,MeV) and $10\,\%$ (75.9\,MeV) to the final
systematic error. Each spectrum has been fit with the sum of two
contributions. A uniform distribution, based on four--body
($ppK^+K^-$) phase space, was used for the background, whereas the
$\phi$ was modelled by its natural line shape, folded with a
Gaussian function ($\sigma\!=\!1\,$MeV/c$^2$) to take into account
the momentum resolutions of the detectors.

\begin{figure}[t]
  \centering
  \vspace*{+1.2mm}
  \includegraphics[clip,width=0.47\textwidth]{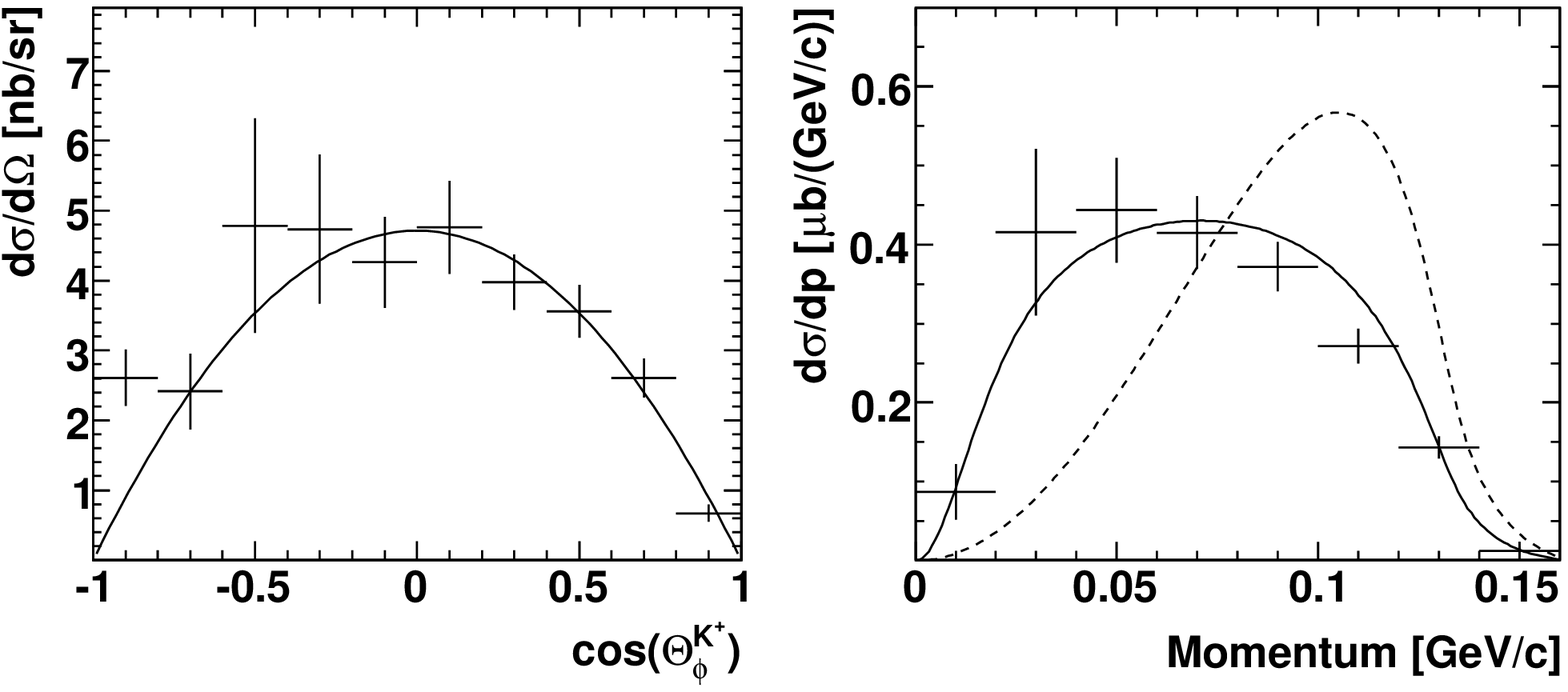}\\
  \includegraphics[clip,width=0.47\textwidth]{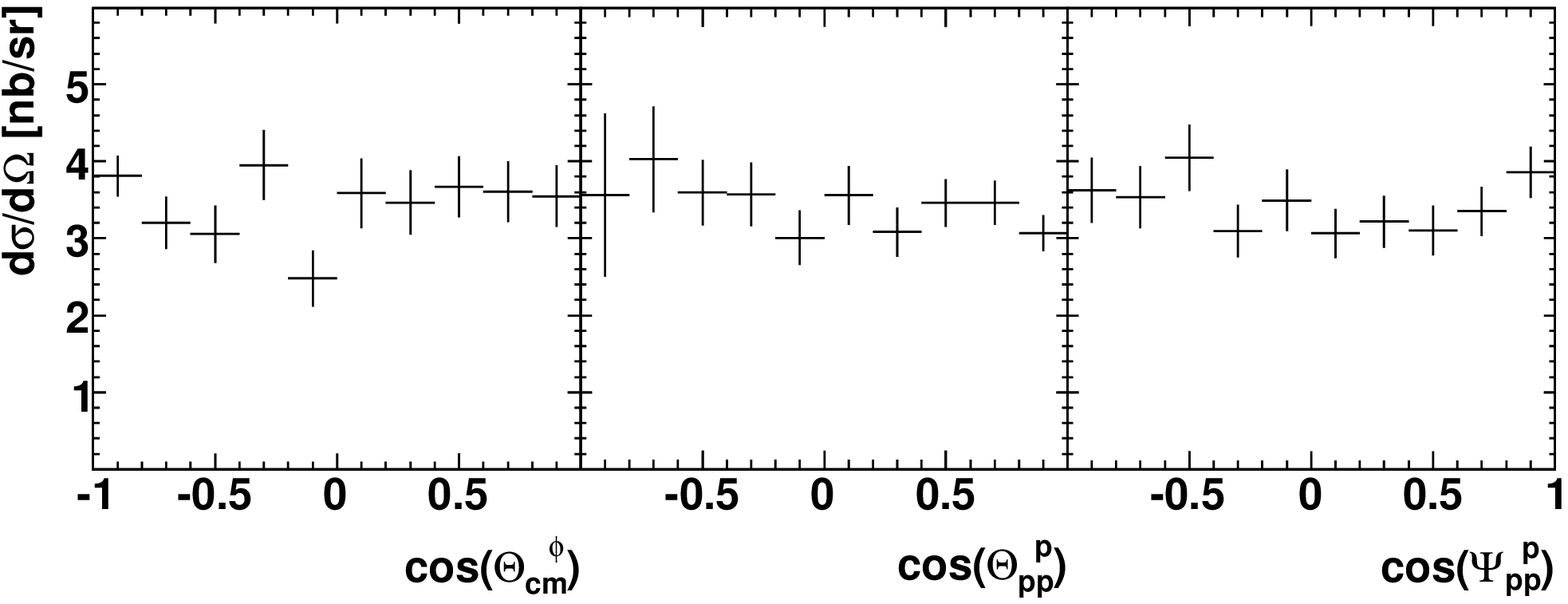}
  \vspace*{-2mm}
  \caption{
Differential distributions for $\epsilon$=18.5\,MeV. Vertical
error bars indicate statistical uncertainties and horizontal ones
bin widths. Upper left panel: d$\sigma$/d$\Omega$ as a function of
the cosine of the polar $K^+$ in the reference frame of the
$\phi$-meson relative to the beam direction. The full line is the
expected $\sin^2\Theta$ shape. Lower panels: d$\sigma$/d$\Omega$
\mbox{\em vs.} cosine of the polar angle of the $\phi$ meson in the overall
\mbox{c.m.} system (left), polar angle of the emitted proton
(middle), and proton polar angle relative to the $\phi$ direction
(right), the two latter being in the ($pp$) reference frame. Upper
right: Dependence of the cross section on the $pp$ relative
momentum. The dotted line reflects pure phase space whereas
the solid includes also the $pp$ FSI.
  }
\label{Diff19MeV}
\vspace*{-2mm}
\end{figure}

Estimates of the differential acceptance of ANKE have been
obtained by means of a multidimensional matrix Monte--Carlo
method, which allows one to determine the acceptance independent
of the ejectile distributions at the production vertex (see
Ref.~\cite{Balestra}). In general, seven degrees of freedom (dof) are
needed to characterize completely the $ppK^+K^-$ final state, but
in our case, a 3--dimensional matrix has been used with the
following dof: (i) the relative momentum of the two final--state
protons in the ($pp$) reference frame, (ii) the polar angle of the
$K^+$--meson in the rest frame of the $K^+K^-$ system, and (iii)
the $K^+K^-$ invariant mass. Significant deviations from pure
phase space can be expected close to threshold for both (i) and
(ii), due to the final state interaction between the two protons,
and also the angular distribution of the decay of the $\phi$
mesons (see Fig.~\ref{Diff19MeV}). The remaining dof are contained
in the implicit assumption of isotropic
angular distributions. These assumptions seem in retrospect to be
justified since the resulting Monte Carlo simulations reproduce
the measured distributions within their statistical uncertainties.
Each of the three variables (i) --- (iii) are subdivided into 10
to 30 bins, producing in total several thousand elements, but for
all of them the acceptance is non--zero. The acceptance
corrections contribute $10\,\%$ (18.5\,MeV), $14\,\%$ (34.5\,MeV)
and $19\,\%$ (75.9\,MeV) to the final systematic error.

\begin{figure}[t]
\vspace*{+0.9mm}
  \centering
  \includegraphics[clip,width=0.42\textwidth]{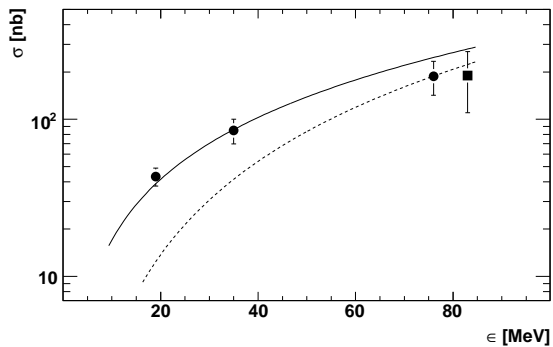}
  \vspace*{-2mm}
  \caption{
Total cross section for $\phi$ production in $pp$ collisions as a
function of excess energy $\epsilon$ from this work (circles) and
DISTO~\cite{Balestra} (square). The error bars include both
statistical and systematic uncertainties. The dashed line shows a
phase space calculation normalized to pass through the highest
energy ANKE point, while the solid line, which includes $pp$ final
state interaction effects, is a fit to all the ANKE data.
  }
\label{TXS}
\vspace*{-2mm}
\end{figure}

Using the number of $\phi$--mesons from the fit, the integral
luminosity for the measurements, and the efficiencies and
acceptances of the ANKE detectors, the total $\phi$--meson
production cross section has been deduced for the three energies,
taking into account the branching ratio in $\phi$ decay of
${\Gamma}_{K^+K^-}/{\Gamma}_\textrm{tot}\!=\!0.491$~\cite{PDG}.
The results are given in Table~\ref{tab:table2} and plotted as a
function of excess energy in Fig.~\ref{TXS}. Very good agreement
is found with the DISTO point at $\epsilon\!=\!83\,$MeV. The
dashed line in the figure displays the energy dependence of phase
space. When this is normalized to the highest energy ANKE point,
it misses the two lower points by large factors. The solid line
includes the effect of the final--state interaction (FSI) between
the two protons in the $^1S_0$ state using the Jost--function
method (see Ref.~\cite{SibBPhi}) and scaled such that it fits best
all three ANKE cross sections. The much improved agreement here
means that it is crucial to include the FSI in any description of the
data.

\begin{table*}
\caption{\label{tab:table2}%
Total production cross section for $pp{\to}pp \phi$ at our three
excess energies (col.~1) compared to $pp{\to}pp \omega$
data~\cite{Hibou,Abd} (col.~2) at similar excess energies. The last column
contains the ratio of $\phi$ to $\omega$ cross sections for each
line. In all cases the first error is statistical and the second
systematic. }
\begin{ruledtabular}
\begin{tabular}{cc|cc|c}
 \multicolumn{2}{c}{$\phi$ production (ANKE)}&
 \multicolumn{2}{c}{$\omega$ production}&
 \multicolumn{1}{c}{$\phi/\omega$ production ratio}\\
 $\epsilon_{\phi}$ [MeV]&$\sigma_{\phi}$(tot)\,[nb]&
 $\epsilon_{\omega}$ [MeV]&$\sigma_{\omega}$(tot)\,[$\mu$b]&
 $R_\mathrm{\phi/\omega}\times10^{-2}$\\ \hline
 18.5&43.2  $\pm$2.2  $\pm$5.1 & 19.6 $\pm$0.9&1.51 $\pm$0.23 $\pm$0.18&2.9 $\pm$0.5 $\pm$0.5\\
 34.5&84.9  $\pm$6.9  $\pm$13.6& 30.0 $\pm$0.9&1.77 $\pm$0.48 $\pm$0.23&4.8 $\pm$1.4 $\pm$0.9\\
 75.9&188.0 $\pm$19.1 $\pm$41.4& 92           &7.5  $\pm$1.9  $\pm$1.5 &2.5 $\pm$0.7 $\pm$0.7\\
\end{tabular}
\end{ruledtabular}
\vspace*{-2mm}
\end{table*}

Before discussing the differential cross sections which were
measured at the lowest excess energy, it is useful to note the
following constraints. Close to threshold, the two final--state
protons must be in the $^{1\!}S_0$ wave, and the $\phi$ in a
relative $S$ wave with respect to this pair, so that the initial
two--proton state is the $^{3\!}P_1$. This in turn requires the
alignment of the incident ($pp$)--spin as well as of the final
$\phi$--meson spin direction to lie along the beam axis (see
Ref.~\cite{Balestra} for a more detailed discussion). The polar
angular distribution of the decay kaons in the $\phi$--meson rest
frame must then display a $\sin^2\Theta_{\phi}^{K^+}$ shape
relative to the beam direction, as is observed for our
$\epsilon$=18.5\,MeV data in Fig.~\ref{Diff19MeV}. Any additional
$\cos^2\Theta_{\phi}^{K^+}$ contribution, induced by higher
partial waves, is not visible. In the lower part of
Fig.~\ref{Diff19MeV}, we show from left to right the distributions
in (i) the polar angle of the $\phi$--meson in the overall
\mbox{c.m.} system, (ii) the polar angle of the emitted protons
relative to the beam, and (iii) the proton polar angle relative to
the $\phi$ direction. Both proton angles are measured in the
($pp$) reference frame. All three distributions are consistent
with isotropy, as expected for a $^3P_1\,{\to}\,^1S_0$ transition.
Finally, in Fig.~\ref{Diff19MeV} the differential cross section is
plotted as a function of the proton momentum in the ($pp$) rest
frame. While the phase--space calculation (dashed line) misses the data,
inclusion of FSI for the two protons in the $^{1}S_{0}$--state
reproduces the experimental results. Thus, a clear and significant
$pp$ final state interaction is observed at
$\epsilon\!=\!18.5\,$MeV.

While at $\epsilon\!=\!83\,$MeV DISTO also observed the dominance
of the $^3P_1\,{\to}\,^1S_0$ transition~\cite{Balestra}, they did
not see any indication of $pp$ FSI in their proton momentum
spectrum, and this is consistent with our findings at
$\epsilon\!=\!75.9\,$MeV. Taking both results together, it is
tempting to ask for the mechanism that suppresses final state
interactions at moderate excess energies: contributions of higher
partial waves are an obvious conventional cause, but more exotic
explanations, like a $\phi N$--resonance (see
Ref.~\cite{SibBPhi}), have also been advanced.

Turning now to the $\phi/\omega$ ratio, we also present in
Table~\ref{tab:table2} data on $pp\,{\to}\,pp\omega$ total cross
sections obtained in the $\epsilon$ range of our
measurement~\cite{Hibou,Abd}. The last column lists the ratios as
obtained line--by--line, \emph{i.e.}\ at slightly different values
of $\epsilon$. Within the stated uncertainties the ratios are
equal, and we have therefore calculated a weighted mean by first
fitting and interpolating the $\omega$ results to our excess
energies. This gives \vspace*{-1mm}
\[ R_\mathrm{\phi/\omega}=(3.3 \pm0.6)\times10^{-2}
\sim 8 \times  R_\mathrm{OZI}, \vspace*{-1mm} \]%
as compared to an uncorrected weighted mean of the last column of
Table I, which is about $10\,\%$ smaller.
Taking into account the effects of the finite meson widths on the 
phase space~\cite{Hibou,SibBPhi} changes $R_\mathrm{OZI}$ by at 
most 5\% at the lowest excess energies.

The production ratio obtained from high energy $ppV$ data is $\sim
(1-2.4) \times R_\mathrm{OZI}$~\cite{LanBoe,Baldi,Blobel}. Together
with our findings, this means that there must be a significant
energy dependence of the OZI enhancement factor~\cite{SibBPhi},
which requires more theoretical work to understand its origin. In
this context let us mention that the experimental ratio
$R_{\phi/\omega}$ deduced from $\pi N$ interaction gives $(3.2
\pm0.8)\times$$R_\mathrm{OZI}$~\cite{SibOZI}, which can be
explained in terms of the established OZI violation in the
$\phi\rho\pi$ and $\omega\rho\pi$ coupling~\cite{GellMann,PDG}.
The present ratio from near--threshold $\phi$ and $\omega$
production in $pp$ collisions exceeds this value by more than a
factor two. It may be a signal for additional, and as yet
non--understood, dynamical effects related to the role of
strangeness in few-nucleon systems.

In summary, we have measured cross sections for $\phi$ production
in $pp$ interactions at three excess energies, all of which are
much closer to threshold than previous data. The lowest energy
result demonstrates the dominance of the transition from the
$^{3\!}P_1$ ($pp$)--entrance channel to the $^{1\!}S_0$ ($pp$)
final--state. Both the total cross section and the
proton--momentum spectrum indicate a significant $pp$ final state
interaction. Using data for $\omega$--production from literature,
it is found that
$R_\mathrm{\phi/\omega}$ is about $8\!\times\!R_\mathrm{OZI}$.

Useful discussions with J.\,Haidenbauer, C.\,Hanhart,
U.--G.\,Mei\ss{}ner, A.\,Sibirtsev, C.\,Wilkin and members of the
ANKE Collaboration are gratefully acknowledged. Our thanks apply
also to the COSY machine crew for their support. This work was supported 
by: BMBF, DFG, Russian Academy of Sciences, and COSY FFE.


\end{document}